# Evaluating an Automated Mediator for Joint Narratives in a Conflict Situation

Massimo Zancanaro[a, b]*, Oliviero Stock[b], Gianluca Schiavo[b], Alessandro Cappelletti[b], Sebastian Gehrmann[c], Daphna Canetti[d], Ohad Shaked[d], Shani Fachter[d], Rachel Yifat[d], Ravit Mimran[d], Patrice L. (Tamar) Weiss[d]

[a]*University of Trento,* [b]*Fondazione Bruno Kessler (FBK), Trento, Italy;* [c]*Harvard University, Cambridge, Massachusetts, United States,* [d]*University of Haifa, Haifa, Israel*

Corresponding author: Massimo Zancanaro massimo.zancanaro@unitn.it – c/o University of Trento – Department of Psychology and Cognitive Science - corso Bettini, 84 I-38068 Rovereto (TN) - Italy

*Word count: 9151*

**Abstract.** Joint narratives are often used in the context of reconciliation interventions for people in social conflict situations, which arise, for example, due to ethnic or religious differences. The interventions aim to encourage a change in attitudes of the participants towards each other. Typically, a human mediator is fundamental for achieving a successful intervention. In this work, we present an automated approach to support remote interactions between pairs of participants as they contribute to a shared story in their own language. A key component is an automated cognitive tutor that guides the participants through a controlled escalation/de-escalation process during the development of a joint narrative. We performed a controlled study comparing a trained human mediator to the automated mediator. The results demonstrate that an automated mediator, although simple at this stage, effectively supports interactions and helps to achieve positive outcomes comparable to those attained by the trained human mediator.

**Introduction**

In social- or ethnic-based conflict situations, mediation is is typically used to heal scars, to reconcile diverse accounts of the past and to build a more positive attitude toward the other side. For example, the intergroup dialogue technique (Allport, 1979; McNamee & Gergen, 1999) is a process in which participants deal with disagreements through self-expression and listening to others. Previous studies used media as a tool to identify solutions to intergroup conflict in order to reduce intergroup stereotypes (e.g., Paluck 2009; Paluck and Green 2009). It has been shown that digital tools supporting intergroup dialogue have beneficial effects on the mediation (e.g., Amichai-Hamburger & Furnham, 2007; Ellis & Maoz, 2007) and that computer-supported cooperative tasks lead to more positive outcomes in terms of reducing prejudiced perceptions and promoting intergroup attraction than did unstructured meetings (Alvídrez et al., 2015).

In social psychology, narration has been related to the construction of one's self-identity (Baumeister and Newman, 1994; Howard, 1991), and many approaches have used narration and storytelling as means for fostering identity recognition and reconciliation of conflicts (e.g., Bar-On & Adwan, 2006; Luwisch, 2001; Maoz et al. 2002; Winslade & Monk, 2001; Hammack, 2008, Zancanaro et al., 2012). Specifically, the framework we adopt in this study has been informed by a cultural-psychology approach (e.g. Hammack 2008; Noor et al., 2012) that suggests an integrative model of identity in which cultural and national narratives (the collective identity/memory) are presented and discussed.

      For structured intergroup meetings, the role of the facilitator or mediator is often crucial to achieving a positive and satisfactory outcome (Winslade & Monk, 2001; Nagda, 2006; Dessel and Rogge 2008). Typically, the mediator is a person with expertise in conflict mediation that provides clear instructions to the participants.

Moreover, the mediator supports the process by ensuring that the participants share their views and are open to compromise.

In this paper, we present the design of *Communics*, a tool that supports the joint creation of narratives by two participants who do not share prowess in the same language, as is typical of many conflicts involving different ethnic groups, minorities or immigrants. The narratives are in the form of illustrated stories on the topic of the Israeli-Palestinian conflict. In contrast to previous work (e.g. Ioannou et al. 2013; Zancanaro et al. 2012), the participants are not co-located and there is no human intervention for translation. The main goal of this paper is to demonstrate the feasibility of automated mediation between remote participants in order to guide them towards an effective interaction. The progress of a story is closely monitored by an automated mediator component that remotely logs the interface events in real-time. Following a number of predefined conditions, the mediator sends messages to the participants to escalate or de-escalate their contributions to address and resolve points of conflict. These interventions are designed to encourage the participants to create and reach an agreement on a joint narrative that is acceptable to both participants, and yet maintains their identities and points of view.

We evaluated the use of this tool with 48 pairs of Hebrew and Arabic speaking young adults who are students at University of Haifa in Israel. The results provide evidence that this tool enables positive changes in attitudes toward the other population. Moreover, an automated mediator is able to monitor exchanges and make timely suggestions regarding contributions that escalate and de-escalate the shared narration in a manner that is comparably effective to the role played by a human mediator.

***The Communics Tool***

The *Communics* tool is a simple interface that allows two participants to remotely create a narrative in the form of an illustrated story (Figure 1). The design of the tool explicitly considers the following assumptions and design goals:

(1) *Two participants.* We focused on a dyadic activity where only two participants contribute to the joint narrative, a situation that in the future can also be expanded to work by teams from two sides or groups.

(2) *Joint but remote work.* Joint activities are crucial for intergroup dialogue (Allport, 1979) but face-to-face encounters can be logistically complicated to organize and manage (see for example, (Ellis and Maoz, 2007; Zancanaro et al., 2012). Online contact has been shown to be, in many cases, a valid alternative to face-to-face encounters (Amichai-Hamburger, 2015).

(3) *Use of own language.* Even in the cases in which both participants share a common language (as in our case where most Arabic-speaking citizens of Israel are relatively fluent in Hebrew) it is important that each individual is allowed to speak their native language to foster a symmetric and non-partisan relationship (Bekerman & Horenczyk, 2004).

(4) *Escalation and de-escalation.* Following Winstok (2008), the tool enables participants to escalate the confrontation in order to express their own points of view before guiding them to de-escalate toward a joint agreement. In a storytelling activity, this means that narratives should not be neutral stories. Rather, a small number of confrontational contributions by both sides should be encouraged before leading the narrative towards a jointly agreed upon conclusion.

In line with the above assumptions and goals, *Communics* is a desktop application that supports two remote participants to collaboratively narrate a story. Participants jointly compose a short illustrated story by taking turns in choosing backgrounds and populating them with characters (each with a combination of body postures and facial expressions), objects and language expressions.

During each turn, a participant can create a new frame in the story by selecting a background image from an image library. In this study, the library contained 33 images of local scenes that were associated with a negative, neutral, or positive sentiment. After selecting the frame, the participants can add two characters to it, which are available in seven different postures. We additionally provided 22 images of everyday objects to enrich the frame with greater context.

The development of the story is driven by the textual narrative and the dialogue between the characters. Participants choose from a library of 157 predefined language expressions to create the story by describing situations and dialogue between the characters. Three domain experts, two from the field of intergroup relations and one from language studies, annotated each language expression with two "sentiment" values, one for each side of the conflict (i.e., the Hebrew and Arabic speakers). The sentiment expresses the tendency of an expression to escalate the conflict (-1), de-escalate the conflict (+1) or maintain a neutral status (0). To ensure validity for annotations, the three experts labelled all expressions separately. Then, all expressions with less than 100% inter-annotator agreement were discussed until the annotators reached agreement.

Selected language expressions are inserted as speech bubbles, similar to those used in illustrated stories or comics. These expressions had been previously translated and aligned in two languages (Arabic and Hebrew). As the two participants are remotely located, the respective *Communics* apps are synchronized to show the same content, with each set shown in the appropriate language for that participant. That is, each participant contributes to and sees the whole story in their native language.

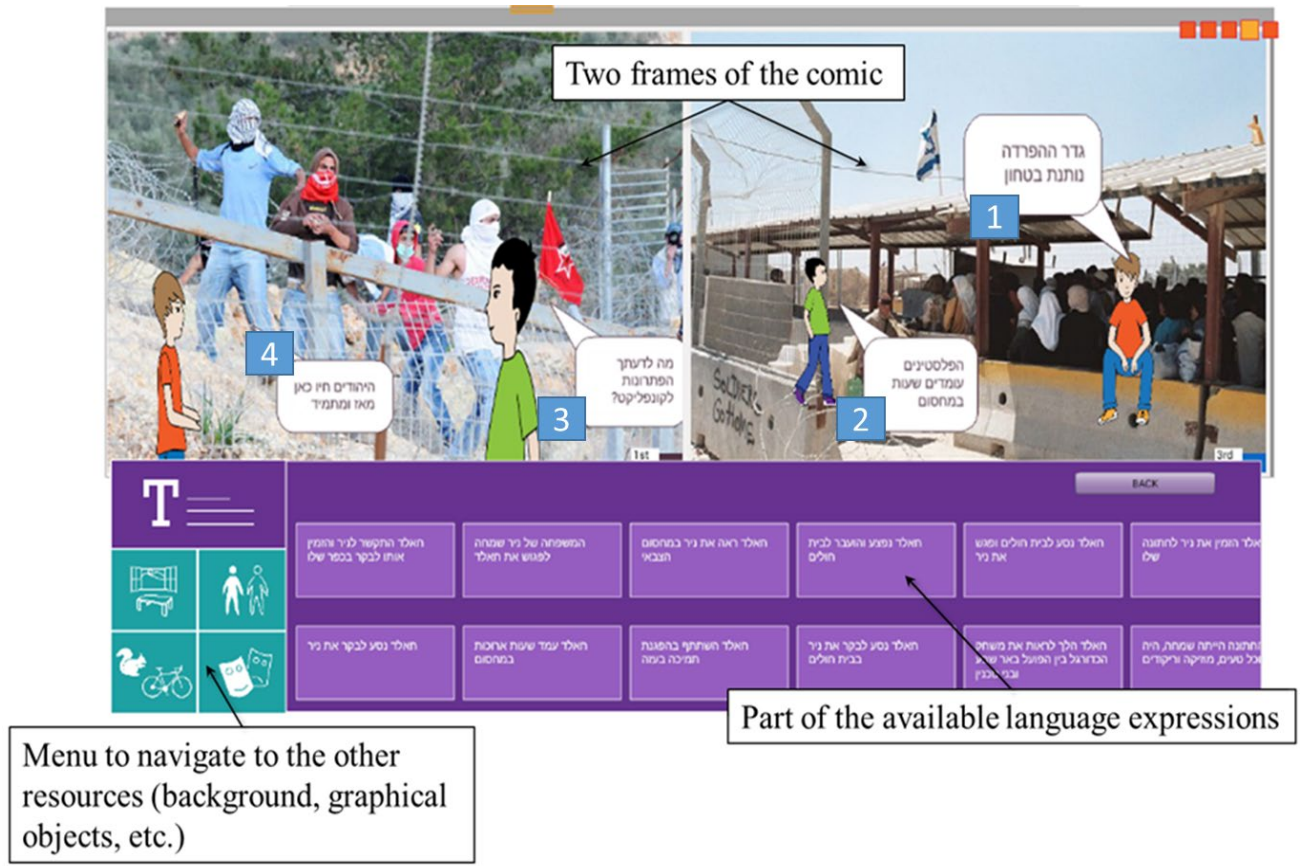

*Figure 1. An annotated screenshot of the Communics interface. It depicts part of a story realized in the study. The balloons are in Hebrew and have the following texts: (1) A border barrier provides security; (2) The Palestinians stand for hours at the barriers; (3) What do you think can serve as solutions for the conflict? (4) The Jews have lived here forever.*

Figure 1 shows a screenshot of the interface as it appears to a Hebrew-speaking participant. The upper half of the interface shows two frames of the current story. The lower half of the interface shows part of the library of language. The buttons on the lower left side let the participant switch between the selection of backgrounds, objects, and characters.

### The Automated Mediator

A mediator's main task within conflict coaching is to listen and ask open-ended questions to engage the participant(s) in developing new ways of thinking about the conflict situation (Brinkert, 2013). In our specific case, the mediator guides the development of the narration, and ensures that story development progresses.

The first reason for including an automated mediator in *Communics* is the scarce availability of human mediators, especially considering that a mediator should have expertise in the conflict, speak both languages and be able to translate them in real time. Additionally, the ethnicity of a mediator may potentially bias the participants (King, 2014; Nagda & Zúñiga, 2003; Zúñiga, Naagda & Sevig, 2002). For example, an Israeli mediator may lead an Arab participant to believe he or she is on the "wrong" side of the conflict, while an Israeli participant may be encouraged to take a more inflexible position due to the apparent partiality of the mediator.

Following the aim described above, the automated mediator takes a passive role in the narration and observes the interaction between the two participants. Following a "heuristic-based" mediation style (Carnevale et al. 1989), it tracks every action within the two interfaces and, when predefined criteria are fulfilled, it intervenes by prompting one or both participants with a context-dependent comment. The design of the criteria follows the *escalation and de-escalation* theory (Winstok, 2008) that posits that interactions among participants should not be kept neutral. Rather, the interactions should escalate to make the conflict apparent to both parties and then de-escalate towards a compromise that both parties agree to. The automated mediator is designed to recognize the patterns of exchanges that may signal that an interaction is proceeding in a neutral, escalating, or de-escalating way. Specifically, the automated mediator aims to control the narration via the following three criteria: (i) the story should escalate so that both parties can express their points of view, (ii) towards the story's end (i.e. after at least one escalation), the participants should de-escalate conflicts (iii) in order to find a mutually acceptable common solution to the story, both participants should have approximately equal contributions to the story. In order to measure the escalation level of a story in progress, the mediator considers the current and the previous levels of escalation based on textual expressions in a story.

Since each textual expression in the library is annotated with a sentiment value for each side of the conflict, the mediator can compute two escalation levels. To derive the escalation level, we first consider the sentiment of a textual expression in isolation, i.e. what effect it would have without considering prior decisions. We refer to this value as the *intrinsic value* (*s*) of an expression. We use the intrinsic value as the baseline effect that adding or deleting a textual expression has on the story escalation. This value is modulated relative to the overall escalation level.

The open-questions, posed by the automated mediator, are aimed to foster the interaction in a desired direction (that is, toward escalation or toward de-escalation). They were prepared by two political psychology doctoral students who specialize in the Israeli-Palestinian conflict. Their previous experience included a number of years in guiding groups of Israelis and Palestinians in both face-to-face and virtual encounters. After helping define the messages, the same doctoral students have been involved as human mediators in the control condition of our study.

We illustrate this with hypothetical participants Alice and Bob. An escalating expression by Alice in a non-escalated (i.e., neutral) context has a larger impact on Bob's *escalation level* than the same expression in a situation with an already elevated *escalation level*; on the other hand, a de-escalating expression has a larger impact in an escalated situation than in an already de-escalated context.

In order to account for deletions of expressions, we consider the immediate impact an interaction has in isolation from the overall impact. Consider an example in which Alice adds an escalating term to the narrative. Bob is offended by seeing this term and creates a response with a similar *intrinsic sentiment value.* If Alice deletes the initial expression, the immediate impact on the frame vanishes. However, the action has a lasting impact on the *escalation level* of Bob. Therefore, the mediator considers every added expression in the context of its preceding actions (and their *intrinsic values*). Further, we assume a hierarchical escalation within a story. Specifically, frames in the context of the story are treated similarly to expressions within the context of a frame. We assume that each frame in a story has a separate *escalation level* (*e*). Similar to the textual expressions, the escalation level of a frame depends on the escalation level of its context. When Alice adds a new frame to the story, the frame's *escalation level* depends on the escalation of previous frames. The immediately preceding frame has the highest impact on the *escalation level;* the farther away a frame is in the current story sequence, the smaller its impact. Therefore, for each new frame added to the story, the automated mediator computes its *initial escalation level* by summing up the *escalation levels* of the previous frames with a discount factor that penalizes the older frames in the story.

$$e_{|n|+1} = \sum_{i=1}^{|n|} e_{n_i} * \gamma^{|n|-i}$$

*Equation 1. Calculation of initial escalation level of a frame*

To concisely describe our strategy as a series of calculations, let a story comprise a set of *n* frames $F_1,...,F_n$ with associated escalation levels $e_1,...,e_n$. In order to control the influence of past actions we further introduce a discount factor γ. The initial escalation level of a newly created frame is described in Equation 1. This formulation allows for an exponential decay of influence with increasing distance from the created frame that is controlled by γ. An exponential decay function has been chosen to account for the fact that antecedent frames should contribute to the escalation level of the current one, in a sharply decreasing manner depending on their distance. The parameter γ has been chosen empirically by observing data from initial pilot studies and set to 0.65.

Next, let *v* be the impact of the insertion of a language expression on the escalation level *e* of its frame (dropping the index for brevity), and *s* be the intrinsic value of the language expression (pre-computed as described above). We subtract a normalizing effect from s that is defined by *e*. As described above, the normalizing effect causes a negative element in a negative frame to have a less influence than a negative element in a positive frame. We achieve this effect by applying a *tanh* to *e* and multiplying it by a second control parameter $\lambda$, as shown in Equation 2.

We chose values $\lambda$ based on a small number of successful pilot-tests with the goal to model the escalation process and set to the actual value of 0.65. While we note that other formulations are possible, the iterative collection of pilot data in such sensitive domains of conflict required for exhaustive evaluation is not feasible.

$$v = s - \lambda * tanh\ e$$

*Equation 2. Calculation of impact of an object*

It is important to reiterate that the only means of communication by the automated mediator is a set of language-based interventions (open questions), which are triggered by the state of the story and the corresponding levels of escalation (whether the frames are escalating or de-escalating). The interventions appear as text messages on the participants' interfaces. Both participants are aware of all the mediator's messages (in their own language), no matter who is addressed in a message. In the future, we plan to experiment with the use of mediator messages that are not necessarily transparent to both participants. At present, the mediator–driven interventions include the following rules and related questions:

**Foster escalation.** A mediation action is triggered if, after four warm-up turns, tno frames above a predefined escalation level threshold have been created (this threshold of value 0.3 has been set following several pilot trials ). The automated mediator sends a message to the participants asking them to explicitly select narrative expressions that reflect their points of view.

**Initiate de-escalation.** If there are two or more turns above a predefined escalation level threshold (again, set to a value of 1 following several pilot trials), the automated mediator sends a message asking the participants to select ideas related to resolution of their conflict.

**Manage viewpoints.** A viewpoint conflict occurs when one of the two participants consistently attributes positive or negative expressions to specifically one of the two characters in the story. In this case, the mediator sends a message asking that participant consider the point of view of the other side. This can be either a trigger for escalation or for de-escalation.

**Manage contribution asymmetries.** The last rule does not use escalation level but simply considers whether there is a difference in the relative number of contributions by the two participants (e.g., number of frames created, and objects and language expressions used). If one participant contributes more than 66% of all actions, the mediator sends a message to foster greater participation to the less prolific participant.

In order to avoid repetitive or too frequent interruptions from the mediator that may eventually irritate the participants, the system never sends the same message more than once in the same turn but waits for at least two turns before sending it again. Moreover, escalation messages are never sent before the third turn and nor after the tenth turn while de-escalation messages are not sent before the eighth turn.

**Experimental questions and approach**

Our main goal was to assess the effect of the automated mediator as a means of effectively supporting the narration task using *Communics* by leading to a more positive attitude in conflict situations. Although there is evidence that cognitive tutors are effective in providing scaffolding in learning situations as presented above, there are no studies concerning the effectiveness of an automated approach to support the mediation between two participants in a joint narration task. Therefore, we formulate our first research question as following:

- *RQ1*: In comparison to a human mediator, are the actions of the automated mediator sufficient in both quality and quantity to motivate the participants toward a satisfactory interaction?

In order to support RQ1, we need to provide evidence about effectiveness of *Communics,* and therefore, formulate an additional research question:

- *RQ2*: Can engagement in a *Communics*-supported joint narration task by participants from two populations in conflict lead to positive attitude changes within and between the groups?

With regard to RQ2, the ability of a joint narration task to induce more positive attitudes toward the other side in a conflict is well-known in the literature as discussed above and did not aim to compare the effectiveness of *Communics* to other approaches, but rather investigate whether this specific tool is effective. We therefore assess the change of attitudes after the experience with *Communics* and compare it between different mediation conditions.

*Method*

Within a 2x2 design in this study, we varied the type of mediator (human/automated) and type of participant (Hebrew speaker/Arabic speaker).

Two trained mediators operated for the *human mediator* condition to remotely observe the interaction between the participant pairs and send appropriate escalation/de-escalation textual messages to them through a chat window. The human mediators were political psychology doctoral students who specialize in the Israeli-Palestinian conflict and have been involved for a number of years in accompanying and guiding groups of Jews and Arabs in both face-to-face and virtual meetings. The team consisted of an additional two assistants who represented the Arabic speaking populations. They were available to provide face-to-face assistance in both conditions and they offered textual messages for mediation in the *human mediator* condition. It was either of the Arabic speaking assistants who came into direct contact with the Arabic speaking participants and either of the Hebrew speaking mediator who came into contact with the Hebrew speaking participants.

Pairs of participants were randomly assigned to either condition. During the session, they did not know whether they were interacting with a human or with an automated mediator. For the *type of participant,* participants were assigned to the condition according to their native language (Arabic or Hebrew).

The role of the mediator is crucial to achieving a positive and satisfactory outcome (Winslade & Monk, 2001; Nagda, 2006; Dessel and Rogge 2008; Maxwell et al, 2012). In particular, it has significance in cases of complex dialogue and high-intensity conflict (Slotte & Hämäläinen, 2015 ; Brinkert, 2013) when in fact they direct the significant cognitive and emotional change that takes place in the dialogue (Zariski, 2010). In the absence of any form of mediation, it is likely that the encounter becomes an unstructured discussion. Indeed, initial pilot studies with *Communics* made apparent that a "no-mediator" experimental condition was not an option since the participants were unable to sustain the interaction and the sessions ended without any narrative result. This result is corroborates previous findings in the field of *Computer-Support Collaborative Learning* wherein free collaboration among peer students does not systematically produce learning (see, for example, Dillenbourg, 2002). Similarly, several unmoderated intergroup encounters via Facebook to promote a reconciliation dialogue between Jewish and Arabic speaking participants resulted in hate filled discussions or were inactive (Ruesch, 2011).

*Participants*

Forty-eight Israeli Arabs (mean age M ± SD =27.0 ± 2.8 years) were paired with 48 Israeli Jews (mean age M ± SD =25.0 ± 3.5 years). All the participants (70 females and 26 males) were students at the University of University of Haifa. Ethical approval was obtained from the Faculty of Social Sciences' Ethical Review Board.

*Measures*

Measures from several sources were gathered prior to, during and following the session.

    *Pre-session measures*. We consider demographic data (age and gender) and the level of *prior contact with the other side* as possible confounding variables. The latter is based on Yuker and Hurley (1987); three items were asked and rated on 6-point scale (1=not at all and 6=very much): 1. Do you have Jewish/Arabic friends? 2. Do you have Jewish/Arabic friends on social media such as Facebook and Instagram? 3. Do you meet with Jewish/Arabic acquaintances frequently at work or school? Higher values indicate a higher frequency of previous contact with the other side.

    *Pre- and post-session measures.* These measures are aimed at assessing the impact of the narration task (research question RQ2 above) and include:

- *Willingness to compromise*. 4-items selected from Halperin and colleagues' questionnaire (2011) aimed at testing participant's attitude toward compromise regarding the Israeli-Palestinian conflict and negotiation (6-point scale ranging from 1 "not at all" to 6 "very much").

- *Willingness to learn more about the other side (the outgroup)* includes three items investigating willingness to (i) hear more about the other partner group's perspective about the conflict, (ii) communicate with them via a social network, and (iii) participate in other online joint activity between Jews and Arabs in Israel. The highest values represent the highest frequency of willingness to learn more about the other side. These scales (6-point scale ranging from 1 "not at all" to 6 "very much") were taken from previous research and adapted for our purposes (Shelton and Richeson, 2005; Halperin et al., 2012).

- *Attitude toward the conflict.* An 8-item questionnaire rated via a 6-point scale developed by Maoz and McCauley (2008), higher scores mean a more positive attitude.

- *Anger toward the other group.* Based on Halperin (2012), participants were asked to rate their feelings of anger toward the other side via a 6-point scale (higher scores mean more anger).

*Post-measures.* These measures were aimed mainly at assessing the participant's experience of the interaction within *Communics* and their perception of the mediator (RQ1).

- *Interaction experience*. Four items on a 6-point scale (1= not at all; 6 = very much) aimed at assessing four different dimensions: *Contact* ("Did you feel that you got to know your partner as if you met him/her in person?"); *Ease of use* ("Do you think that this app is easy to operate technically?"); *Satisfaction* *regarding the interaction* ("Do you feel satisfied with the way the interaction went with your partner?"); *Satisfaction regarding the story* ("Do you feel satisfied with the story created with your partner?"); *Expressiveness* ("Do you feel that the experience enabled you and your partner to express yourselves in the best way?"). These items were adapted from previous studies (Dennis and Kinney, 1998; Hecht, 1978; McGloin et al. 2011; Warkentin and Beranek, 1999). For each item, higher values reflect a more positive experience.

*Participant's perception of the mediator*. Five items rated using a 6-point scale (1 = not at all; 6 = very much) measure the perception about the mediator's role during the session. Following Zarinski (2010), the function of dialogic mediation is to stimulate effective and efficient problem solving and decision making by the participants. We believe it is ethically acceptable that the means used by the mediator be transparent, and that perceived effectiveness and efficiency are important dimensions to measure: *Effectiveness* ("Did the mediator guide you to build a balanced story?"), *Efficiency* ("Did the mediator intervene too often?", a reversed item). Related to the previous, preliminary questions are an additional two. Fairness and trustworthiness are recognized to be key characteristics of a successful mediator (Arad & Carnevale, 1992); we measure them with the items, *Trustworthiness* ("Did you trust the mediator?") and *Fairness* ("Was the mediator more likely to favour the other participant?"). Finally, for intercultural encounters, balanced participation is regarded as a positive characteristic: *Balanced participation* ("Did your partner contribute more than you to constructing the story?").

Furthermore, the participants responded to a binary (yes-no) question concerning the mediator's identity: "Did you think that the mediator was a computer and not a human?"

*Session log measures. Communics* usage logs were recorded for each session. These data included the duration, the number and type of text phrases (categories of texts and levels of escalation) used in total during each session and as used by each of the participant pairs, number of items created and number of turns taken.

### *Experimental Hypothesis*

Regarding RQ1, we hypothesized:

- H1a: The *interaction experience* measures will not differ significantly with respect to the type of mediator for either group.
- H1b The *participant's perception of the mediator* will not differ significantly with respect to the type of mediator for either group.

Regarding RQ2, we hypothesized:

- H2: The measures Willingness to compromise, Willingness to learn more about the other side, Attitude toward the conflict and Anger toward the other group will improve significantly from the pre-test to the post-test for both groups and for both mediator conditions.

### *Procedures*

Three days before the experiment, participants received the pre-questionnaire. On the day of the experiment, participants were randomly assigned to a partner from the other group. The pair was randomly assigned to one of two conditions, human or automated mediator. Participants were instructed in the way they were expected to perform the task, i.e., create a joint story, with each one of the pair taking turns to select graphic backgrounds, characters, objects and text expressions, and to respond to their partner's selections. While instructing the participants, we were carefully to avoid any bias with respect to the mediator being a human or a computer. Furthermore, we ensured that participants had equal opportunities to express themselves as well to contribute to the progress of the story.

The participants were located in separate rooms, with an assistant who spoke their language and provided instruction and technical support; they did not physically meet each other prior to, during or following the session. The human mediator interacted with the participants remotely. Each participant viewed the *Communics* interface on a laptop with a 14-inch screen. After the completion of the joint story, participants were asked to individually complete the post questionnaire items.

The duration of the entire session was about 1 hour including training (10 min), the story-telling task (40 min) and the post questionnaires (10 min).

**Results**

*Pre-test measures.* Table 1 shows the descriptive statistics of the participants' pre-test measures listed by mediator condition and group. No statistical differences were observed between groups with respect to these measures.

*Table 1. Descriptive statistics and results of nonparametric tests (two-sample K–S test) comparing mediator conditions*

| Dimension | | Human mediator | | Automated mediator | | *p* value |
|---|---|---|---|---|---|---|
| | | *Hebrew speakers* | *Arabic speakers* | *Hebrew speakers* | *Arabic speakers* | |
| *Gender* | M | 7 | 6 | 7 | 6 | |
| | F | 18 | 17 | 18 | 17 | |
| *Age (in years)* | | 25.5 (3.96) | 23.05 (3.72) | 23.87 (3.03) | 22.22 (1.95) | .27 |
| *Prior contact with the other side (3-items on a 6-point scale where higher scores mean more contacts)* | | 3.29 (0.99) | 4.26 (1.03) | 3.50 (1.01) | 3.68 (1.4) | .91 |
| *Attitude toward the conflict (8-items on a 6-point scale where,* | | 3.19 (1.00) | 3.57 (0.75) | 2.71 (0.85) | 3.58 (0.77) | .50 |

| | | | | | |
|---|---|---|---|---|---|
| *higher scores mean better attitude)* | | | | | |

*Characteristics of the sessions.* There were no significant differences in storytelling duration between the ones mediated by the computer versus those mediated by a human (*automated mediator* M ± SD=40.8 ± 7.77 min; *human mediator* M ± SD=43.8 ± 7.55 min; $t_{46}$=-1.35, *p*=.184). Neither were there significant differences between the mediator types for any of the main session characteristics including the number of different texts selected by the participants (*automated mediator* M ± 132.3 ± 52.3); *human mediator* M ± SD=135.6 ± 51.2); $t_{94}$=-.32, *p*=.75), the number of turns taken (*automated mediator* M ± =11.9 ± 3.8; *human mediator* M=12.3 ± 3.9; $t_{94}$=-.56, *p*=.57), and the number of items created (*automated mediator* M ± SD=4.4 ± 2.6; *human mediator* M ± =4.4 ± 2.3; $t_{94}$=.01, *p*=.99).

*Analysis of stories.* To evaluate the final narratives, we annotated whether the narratives included: (i) Dialog or Narration statements, i.e., direct speech produced by one of the characters (e.g., "I invite you to a meal at my home.") or General statements (e.g., "They ate ice-cream"); (ii) Social or Political topics, i.e. the stance participants chose to present when communicating with their partner, e.g. "I play football" (Social) or "Soldiers enter the homes of Palestinians" (Political); (iii) Contingent or non-contingent, i.e., the content of a statement produced by one participant was or was not related to the statement produced by the other participant.

Descriptive findings from 48 sessions, 24 with a human mediator and 24 with the automated mediator, revealed that the great majority (86%) of the statements were of a Dialog type. No differences were found between sessions with the automated mediator and the human mediator in terms of the topics used by the participants: with the automated mediator, 58% of the statements were about Social topics and 42% were about Political topics, while with the human mediator, 55% of the statements were about Social topics and 45% were about Political topics. Regarding contingency, the majority of the statements (73%) were related in content to each other reflecting that the level of interactivity between the participants was high.

In summary, creating a joint story was characterized by two major choices made by the participants: 1. a symmetry between political and social contents; 2. a clear preference to create a dialogue between the characters in the story, a choice which reflects the influence of comics genres.

*Pre-post measures.* A repeated measures ANOVA was used to test the hypothesis concerning a significant effect of responses prior to and following a single session with the application, including *type of participant* and *type of mediator* as between subject factors. Overall, the analysis showed a significant multivariate effect of the within-subject *time* (pre vs. post tests) factor ($F_{4,88}=21.71$, $p<.01$, partial $\eta^2=0.5$), an interaction between *time* and *type of participant* ($F_{4,88}=4.12$, $p<.01$, partial $\eta^2=0.16$), and a significant effect of the *type of participant* ($F_{4,88}=5.8$, $p<.01$, partial $\eta^2=0.22$). No differences between *mediation types* were observed. The univariate tests showed a difference between pre- and post-tests for all the measures (*Attitude toward the conflict*, *Willingness to learn more about the other side*, *Anger toward the other,* all *p<.01*) except for *Willingness to compromise* (*p*= .17). A significant interaction emerged for *Anger toward the other (p<.01)* and *Willingness to learn more about the other side* scales (*p<.05*) indicating a stronger decrease in *Anger toward the other* scores for the Hebrew-speaking group and a higher *Willingness to learn more about the other side* for the Arab-speaking participants (Table 2).

To better interpret the results, an equivalence hypothesis testing approach, the "Two-One-Sided t-test" (TOST) procedure, was used to test whether the means of post-test measures for the two groups are close enough to be considered equivalent (Lakens, 2017). In equivalence tests, the null-hypothesis is that there is an effect, whereas the alternative hypothesis tests whether the observed effect is smaller than a specific value, referred to as the Smallest Effect Size of Interest (SESOI). In our analysis we set Cohen's d to the value of 0.6 (indicating a medium effect) as our SESOI, following the approach suggested by Lakens et al. (2018). The test showed that all post-session dimensions were equivalent between the two conditions, with the exception of the *attitude toward the conflict* (Table 2).

*Post-session measures.* A MANOVA was used to compare the scores between the types of mediation. For the interaction experience, there was no significant multivariate effect ($F_{5,86}= 3.1$, $p = .06$). Considering the results of the equivalence testing, we can conclude that the differences between all the items, except for *Contact*, were equivalent.

Table 2. Mean and Standard Deviation for questionnaire items on 6-point scales (the higher the better). Equivalence tests are based on Welch's t-test and use TOST procedure with d= 0.6 and ⍺=0.05. For the intervention effect scale, only post-session measures were considered in the tests.

| Scale | Dimension | Human mediator | Automated mediator | Equivalence test |
|---|---|---|---|---|
| **Intervention effect** | *Willingness to compromise (4-items on a 6-point scale where higher scores mean more willingness to compromise)* | Pre: 3.64 (1.32)<br>Post: 3.60 (1.1) | Pre: 3.60 (1.2)<br>Post: 3.41 (1.28) | $t(91.9)=-2.16$, $p= .02$<br>[90% CI= -0.72, 0.59] |
| | *Willingness to learn more about the other side (3-items on a 6-point scale where higher scores mean more willingness to learn)* | Pre: 3.48 (1.13)<br>Post: 3.76 (1.43) | Pre: 3.38 (1.12)<br>Post: 3.56 (1.2) | $t(91.2)=-2.19$, $p= .02$<br>[90% CI= -0.25, 0.65] |
| | *Attitude toward the conflict (8-items on a, 6-point scale where higher scores mean better attitude)* | Pre: 3.37 (0.9)<br>Post: 3.78 (0.89) | Pre: 3.16 (0.91)<br>Post: 3.43 (0.95) | **n.s.** $t(93.6)=1.08$, $p= .14$<br>[90% CI= -0.04, 0.66] |
| | *Anger toward the other (6-item, on a 5-point scale where higher scores mean more anger)* | Pre: 3.04 (1.21)<br>Post: 2.18 (1.3) | Pre: 3.38 (1.12)<br>Post: 2.40 (1.26) | $t(93.9)=2.10$, $p= .02$<br>[90% CI= -0.65, 0.21] |
| **Interaction experience** (4-items on a 6-point scale where higher scores mean better experience) | Contact | 2.75 (1.24) | 3.32 (1.1) | **n.s.** t(92.7)=0.56, *p*= .29 [90% CI= -0.97, 0.17] |
| | Ease of use | 3.75 (1.08) | 3.84 (1.22) | t(92.6)=2.56, *p*= .01 [90% CI= -0.48, 0.31] |
| | Satisfaction with interaction | 3.84 (1.31) | 3.88 (1.12) | t(91.8)=2.77, *p*= .01 [90% CI= -0.45, 0.37] |
| | Satisfaction with story | 3.93 (1.21) | 3.78 (1.22) | t(93.9)=-2.33, *p*= .01 [90% CI= -0.26, 0.56] |
| | Expressiveness | 2.91 (1.49) | 3.16 (1.39) | t(93.5)=2.09, *p*= .02 [90% CI= -0.73, 0.24] |
| **Mediator perception** (5-items on a 5-point scale where, higher scores mean better perception of the mediator) | Effectiveness | 3.78 (1.60) | 3.58 (1.60) | t(94)=-2.33, *p*= .01 [90% CI= -0.34, 0.74] |
| | Efficiency | 3.22 (1.2) | 3.42 (1.04) | t(92.1)=2.07, *p*= .02 [90% CI= -0.45, 0.37] |
| | Trustworthiness | 5.15 (1.01) | 4.44 (1.36) | **n.s.** t(92.1)=-0.03, *p*= .46 [90% CI= 0.30, 1.12] |
| | Fairness | 1.46 (1.11) | 1.40 (0.81) | t(86)=-2.64, *p*= .01 [90% CI= -0.27, 0.39] |
| | Balanced Participation | 3.61 (0.99) | 3.25 (0.95) | **n.s.** t(93.8)=-1.12, *p*= .13 [90% CI= -0.03, 0.69] |

Figure 2 shows the results from the *Participant's perception of the mediator* items. The multivariate effect was significant for the mediator type ($F_{5,86}$= 2.17, *p* <.05, partial $\eta^2$=0.15) indicating a difference in the participants' perception of the mediator between the two conditions. The univariate test showed a significant difference in the trustworthiness item only ($F_{1,94}$=8.38, *p*<.05, partial $\eta^2$=0.08). Participants gave significantly higher scores of trustworthiness to the human mediator with respect to the automated one. However, absolute values were relatively high in both conditions. Univariate tests for the other dimensions were not significant (Effectiveness: *p*= .54, Efficiency: *p*= .11, Fairness: *p*=.77, Balanced Participation: *p*= .08). Participants rated the effectiveness and efficiency of the system as relatively high, and perceived that the relative contributions were balanced in both conditions (Table 2). Equivalence testing showed that effectiveness, efficiency and fairness can be considered equivalent between the two mediation conditions.

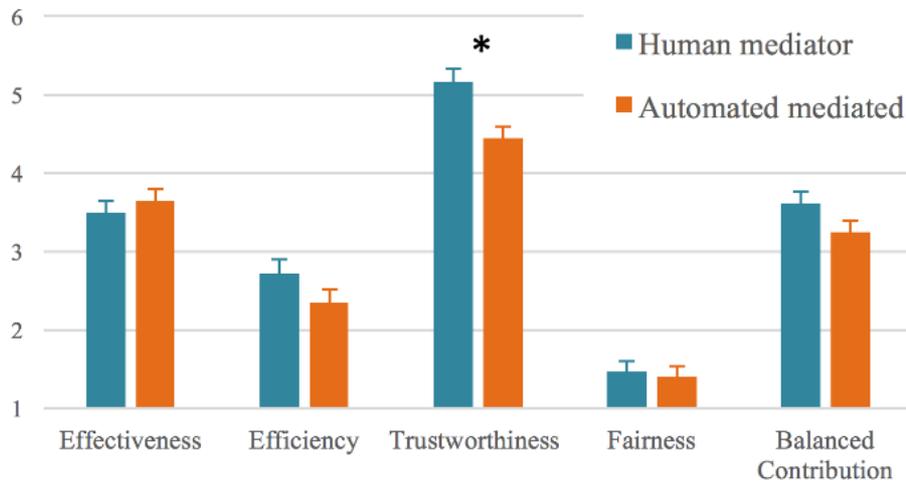

*Figure 2. Scores for the participant's perception of the mediator items. (*only Trustworthiness was statistically different between the two groups)*

To examine factors that contributed to the storytelling outcome, a multiple regression equation was calculated using scores on the *Interaction experience* and *Mediator perception* scales to predict the participants' satisfaction with the story created. As shown in Table 3, the overall model was significant ($F_{3, 97} = 39.29$, $p < 0.01$, $R^2 = .56$). The regression equation found that satisfaction with the interaction, communication expressiveness, and trustworthiness in the mediator predicted satisfaction with the overall story. Thus, higher scores on these scales predict more satisfactory stories.

*Table 3. Linear model of predictors of the satisfaction with the story*

|  | b | SE b | ® | p |
|---|---|---|---|---|
| *Constant* | .53 | .37 |  |  |
| *Satisfaction with interaction (Interaction exp.)* | .48 | .09 | .48 | $p = .001$ |
| *Expressiveness (Interaction exp.)* | .21 | .07 | .26 | $p = .004$ |
| *Trustworthiness (Mediator perception)* | .16 | .07 | .17 | $p = .023$ |

Lastly, when asked after the end of the session, whether the mediator was perceived by the participants as a human or a computer, the participants generally considered the mediator to be a person rather than a computer, even in the automated mediator condition (Table 4). A Chi-square test did not find an association between the type of mediator and the participant perception ($\chi^2=3.6$; $p >.05$).

*Table 4. Participants' perception of mediator being a human or a computer*

|  | I think the mediator was… | |
|---|---|---|
|  | *a computer* | *a person* |
| *Automated-mediator* | 12 | 36 |
| *Human mediator* | 6 | 42 |

**Discussion**

The study results show the potential of using an automated mediator to achieve a positive effect via remote support for pairs of participants on two sides of a political conflict (research question RQ2, hypothesis H2). Specifically, we measured a more positive attitude toward the conflict, an increased willingness to learn more about the other side and a reduction of anger toward the other group.

*Table 5. Summary of the main outcomes of the study*

| Category | Dimension | Main Results |
|---|---|---|
| **Intervention effect** | Willingness to compromise | No difference after sessions for both types of mediator. |
|  | Willingness to learn more about the other side | Increase after sessions, mostly for Arab speaking participants. |
|  | Attitude toward the conflict. | More positive after the sessions. No differences between groups. |
|  | Anger toward the other | Decrease after sessions, mostly for Hebrew speaking participants. |
| **Interaction experience** | Contact | No multivariate effect between conditions. |
|  | Ease of use |  |
|  | Satisfaction with interaction |  |
|  | Satisfaction with story |  |
|  | Expressiveness |  |
| **Mediator perception** | Effectiveness | No differences between conditions. |
|  | Efficiency | No differences between conditions. |
|  | Trustworthiness | Higher scores for human-mediated condition. |
|  | Fairness | No differences between conditions. |

|  | Balanced Participation | No differences between conditions. |
|--|--|--|

Consistent with other short-term interventions (see, for example, the discussion by Alvídrez and colleagues, 2015), we did not find a significant effect on the *willingness to compromise* which is an attitude related to strong and stable political beliefs by people who are involved in violent conflicts and whom have many psychological barriers that impede change (Bar-Tal, 2013; Hameiri et al., 2016), particularly on the basis of a one-time intervention such as (Maoz, 2011) such as used in the current study. Nevertheless, we found positive outcomes for other cognitive and affective measures, which show the potential of our approach. Future studies will examine changes that occur after repeated exposure to the intervention.

Some measures improved only for one of the two groups (Arabic- or Hebrew-speakers). Specifically, only the Arabic-speaking group significantly increased their *willingness to learn* more about the other side and only the Hebrew-speaking participants had a more positive attitude toward the other. This latter result is consistent with the literature which shows that a meeting and ensuing dialogue have more positive effects on emotional and attitudinal change on the stronger, majority group (Bruneau and Saxe 2012; Pettigrew and Tropp 2008; Tropp & Pettigrew, 2005). Future research will aim to better understand these diverging effects, both for the human and automated mediator conditions. We intend to continue to explore explanations for, and the extent to which each side responds differently to the task.

With regard to the ability of the participants to create coherent narratives, the results of the narrative analysis suggest that collaborative narrations through *Communics* are balanced with respect to the political perspective; since political statements are usually used to escalate a conflict while social statements are useful in de-escalation, the achievement of this balance is important. Moreover, the regression analysis showed that three main factors predict satisfaction with the story, namely satisfaction with the interaction with the other partner, expressiveness in the communication, and trustworthiness in the mediator. This finding highlights the importance of interaction with a partner as a key element for creating a satisfying story, and that being able to fully express one's self and having trust in the mediator further contribute to reaching a gratifying narrative outcome.

Regarding the main research questions on the impact of the type of mediation (RQ1), human vs. automated, both hypotheses H1a and H1b were partially confirmed suggesting that an automated mediator appears to be an effective alternative to an expert human mediator when logistical, financial or other reasons warrant it. In particular, with regard to the quality of the experience (H1a), all the items (except *contact*) used to assess the interaction experience (see Table 4) did not differ significantly with respect to the type of mediation. In both conditions, participants expressed a positive level of satisfaction with the ease-of-use of the application, with the way they interacted with their partner, and with the story created. They also felt that the experience, as mediated in both conditions, enabled them to express themselves to an adequate degree.

Similarly, with regard to the users' perception of the mediator (H1b), the scores were relatively high for the automated mediator, and it was regarded as being equally effective and efficient and as fair as the human mediator. We found a small but significant difference for *trustworthiness* which may be related to the content of the messages created by the human mediator which were perceived as more natural than the messages from the automated mediator. Further studies are needed to better investigate this issue.

Finally, it is interesting to note that when asked, at the end of the sessions, whether they perceived the mediator to be a human or a computer, they tended to respond that it was a human, irrespective of the actual condition; more importantly, this perception had no effect on either the attitude or satisfaction ratings.

Overall, this study demonstrated the feasibility of a new type of cognitive tutor aimed not at scaffolding a difficult learning experience but at facilitating the interaction between two participants engaged in a task that is laden with a complex socio-emotional relationship. Although the constrained text, used to facilitate the multi-lingual task, allowed a simpler implementation of our automated mediator, the information used by the mediator is just the sentiment expressed in each contribution and this is already achievable by state of the art in the field of natural language processing (Miner et al., 2012). Therefore, in cases of single language use, it might be possible to extend our approach to free language narration without too many difficulties since the system would not have to cope with the challenge of simultaneous translation.

It may be argued that a limitation of this study is that the expressiveness and quality of the stories may have been constrained by the relatively sparse textual content that was provided to the participants. In preparing the material, the language expressions were carefully selected to provide opportunities for self-expression and, at the same time, were kept to a manageable number (157) in order to facilitate the exploration and accessibility of the library. Although additional work is needed to understand how to carefully calibrate the potential and limitations of our approach, the study results presented in this paper suggest that this structured method is both promising and efficient).

**Related Work**

This field of study brings together different research streams from AI and HCI domains, combining cognitive tutors and storytelling assistants in a technology designed for conflict management and resolution.

*Cognitive Tutors*

The *Communics* automated mediator can be classified in the realm of *intelligent tutoring systems* or *cognitive tutors* which have a long history in the field of Artificial Intelligence (see among others Sleeman and Brown, 1982; Anderson et al. 1995; Oviatt, 2013). They are software programs that guide learners through each step of a problem solution by providing hints and feedback. They have been traditionally designed to help students (generally young learners) acquire skills in different subjects such as algebra, geometry, computer programming and physics. Usually, they include an explicit model of learning which is used to decide when to intervene and which kind of support to offer (Koedinger and Aleven, 2007). Recently, they are becoming mainstream thanks to the success of MOOCs (Massive Online Open Courses; for a discussion see Koedinger and Aleven, 2016).

In general, cognitive tutors are designed to support single users by providing adequate scaffolding in learning tasks (Belland et al. 2017). A meta-analytical review by Van Lehn (2011) confirmed the evidence that automated tutoring for scaffolding appears to be almost as effective as human-based tutoring.

In some cases, cognitive tutors are used to provide explicit support to the collaboration among learning peers. For example, Diziol and colleagues (2011) discussed an approach similar to the current study (although for a scenario of learning mathematics) where the cognitive tutor fosters peer interaction rather than scaffolding learning. In the field of Computer-Support Collaborative Learning (CSCL), the notion of "script" is widely used to represent the tasks that peer learners have to perform as well as the collaboration patterns that occur (Tchounikine et al., 2010). In this context, Fischer and colleagues (2013) discussed an approach of automated support for peers by means of explicit guidance scripts.

The *Communics* automated mediator addresses a novel approach with respect to cognitive tutoring in education. Instead of helping to teach competencies or skills to an individual or groups of students, it facilitates a two-party interaction with the aim of providing a successful shifting of the participant attitudes to each other. As for other cognitive tutors, the *Communics mediator* algorithm was developed around an explicit model that in our case is based on conflict escalation/de-escalation for delivering context-sensitive hints and suggestions to guide users in their joint activity.

***Automated mediation in negotiations***

There have been several attempts to build automated tools to assists parties in conducting a negotiation. Negotiation support systems (Kersten & Lai, 2007) are tools aimed at structuring interaction between two parties and facilitating them to reach an agreement. For example, the Negoisst offers document management tools as well as communication and decision support in a business-to-business e-commerce transactions (Schoop et al., 2003). These type of systems have also been proposed to be applied in conflict management  (Dannenmann & Schoop, 2011). Similarly, the Online Dispute Resolution systems are used to help two parties to structure dispute resolution processes (Bonnet et al., 2003). Usually, these tools require an explicit intention to enter and resolve a dispute and to be able to properly formulate it. Yet, recently several intelligent systems have been proposed to facilitate this process by allowing the participants to naturally express their preferences on multiple, competing issues and without assigning numeric values  (e.g., Chalamish et al., 2012).  A patent has also been also granted related to automated cross-cultural conflict management in which an automated system engages participants in structured conversations and proposes agreements that consider the full ratings of the factors (Femenia & Pomerance, 2004).

The approach taken in *Communics* is different because it does not aim to achieve explicit negotiation but rather to explore cultural conflict; therefore the participants do not have clear and well formulated goals to satisfy during the negotiation. The storytelling approach taken in *Communics* is aimed at facilitating the participants to express and mediate their identity in a conflictual context while the mediator never tries to propose agreements. Nevertheless, some of the techniques used in these tools might be effectively used in further extension of our mediator.

*Digital and collaborative storytelling*

Collaborative storytelling has long been an educational activity to support self-reflection and mutual understanding. In recent years, several attempts have been made to design technology-based devices to assist the narrative process, especially involving group of users during digital storytelling (Göttel, 2011). For instance, Benford and colleagues (2000) designed tablet interfaces to encourage collaboration among children and to invite them to discover the added benefits of working together when creating a common story. Other studies have introduced the use of 2D/3D virtual online environments (Garzotto and Forfori, 2006) that allow remote co-authoring, intelligent control tools (Young and Riedl, 2003) to enrich the narrative experience and conversational storytelling (Chi and Lieberman, 2011). Other works have also explored the use of digital comics for for supporting storytelling activities (Mencarini et al., 2015).

As mentioned above, *Communics* was designed to foster interaction by pairs of peers with the objective of supporting collaborative storytelling in a multi-cultural and intergroup setting. The benefits of adopting collaborative and co-located storytelling for cross-cultural mediation and conflict resolution have been investigated in the HCI domain. For example, studies on shared interactive spaces have shown that collaborative storytelling can be used to facilitate conflict resolution and to support mutual understanding between people with different cultures (Ioannou and Chrystalla, 2016; Zancanaro et al., 2012). These studies investigated the impact of technology-supported interventions for mitigating intergroup conflicts (Stock et al., 2008; Zancanaro et al., 2012), discussing problematic and controversial topics (Ioannou et al. 2013; Ioannou and Chrystalla, 2016), learning about conflict management (Brynen and Milante, 2013) and reflecting on controversial historical events (Pollack and Kolikant, 2012).

In line with this research, the current work further explores the role that interactive technologies can play in mediating intergroup conflict of divided communities and in promoting a greater mutual understanding.

*Online and virtual intergroup encounters*

A virtual meeting has the advantage of overcoming practical and logistic problems that arise when attempting to interact face-to-face. Online technology has a demonstrated potential in promoting intergroup dialogue and favouring accessibility to populations that otherwise would not have any interaction with each other (Amichai-Hamburger and Furnham, 2007; Ellis and Maoz, 2007; White et al., 2015).

This is even more relevant when considering political, and sometimes violent conflicts in which the difficulties in managing co-located interactions become crucial due to the lack of institutional and social support, the high levels of inequality perceived between groups, and the significant reluctance of participants to take part in an on-site joint activity (Ellis and Maoz, 2006; Mor et al., 2016). Furthermore, online encounters may allow participants to maintain anonymity (Amichai-Hamburger and Furnham, 2007), to strengthen their sense of control over an interaction (McKenna et al., 2002), to reduce intergroup bias (White et al., 2015), and to create a more realistic alternative reality which can encourage participation. In contrast, one apparent disadvantage of digital systems versus a physical encounter is in a reduced sense of authenticity (Amichai-Hamburger and Furnham, 2007).

*Communics* is a tool for a specific case of online intergroup encounters. In contrast to other work which used common social networks and their online communities (e.g. Facebook), *Communics* is a software application that supports remote storytelling by tailoring the content to a specific target group (e.g., Hebrew and Arabic speaking young adults living in Israel) and allowing them to communicate in their own language.

**Conclusion**

In this paper, we presented a tool, *Communics,* which supports pairs of remotely located participants to contribute to a shared story, each in their native language. The technology is proposed to support intergroup encounters aimed at fostering reconciliation in social and ethnic conflicts. A key component of our system is the automated mediator that provides a feasible alternative to human interventions during the narration process.

The results of a study involving 96 participants suggest that *Communics* elicits narrations that show a rich and balanced interactivity between the participant pairs. Furthermore, this type of computer-mediated encounter appears to have a positive effect on affective measures, such as decreasing the anger felt toward the other group. Although the current implementation was relatively simple, our study suggests that the automated mediator was perceived as effective and fair as was the control condition expert human mediator (albeit with somewhat less trustworthiness).

*Communics* may potentially for being applied on a larger scale because it supports customization of the material in the library and it does not constrain interaction to be on any specific topic. Similarly, the automated mediator is based on general assumptions and does not require a specific knowledge of the topic under discussion but only to have information about the positive or negative values of each sentiment. The escalation/de-escalation strategy can thus be applied to different contexts in which a conflict takes place.

One key aspect of our study was related to the anonymity of participants, needed because of the sensitivity of this particular conflict. This was preserved in our study mainly because the participants were only anonymous to their storytelling partner but they were always with someone from the research team which may have attenuated circumstances. Yet, the preservation of anonymity may become problematic in different settings, especially large scale applications because it is well known that this is a key factor in antisocial online behavior (Suler, 2004). We believe that the structured process of narration, in contrast to unstructured discussions may help to limit antisocial behavior (Amichai-Hamburger et al. 2015) and automated mediation (although more complicated than the present one) may be leveraged to better support collaboration.

One problematic aspect of our approach is that key aspect of the automated mediation (namely the escalation and de-escalation formulas, the setting of their parameters and the thresholds for their application) have been built out of knowledge of human practices and expert intuition while a data-driven approach would be more robust and reliable. We acknowledge this limitation but a data-driven approach would require a large amount of data that is quite difficult to collect with these types of study. We contend that, due to these initial positive results, future experimentation with an approach of self-regulating settings of the parameters will help to adjust the formulae to different contexts and situations.

A possible limitation of our study is the lack of a no-mediator condition. We note that these types of intervention are inherently stressful for the participants; since our pilot study clearly made apparent that a no-mediator would not have worked, we concluded that it would not be ethically acceptable to expose our participants to a possibly ineffective stressful condition. Indeed, it is quite common, when investigating different mediation approaches, to compare them directly without the no-mediator condition.

Another limitation is the lack of qualitative analysis specifically on the feedback from the participants. Actually, for logistical reasons, we were not able to collect them and in future studies, we will remedy this aspect.

Furthermore, future work will be directed at systematically investigating the implementation of richer mediation strategies as well as assessing this approach to different types of conflict settings.